\documentclass[a4paper,twoside]{article}

\usepackage{epsfig}
\usepackage{subfigure}
\usepackage{calc}
\usepackage{amssymb}
\usepackage{amstext}
\usepackage{amsmath}
\usepackage{amsthm}
\usepackage{multicol}
\usepackage{pslatex}
\usepackage{apalike}

\usepackage{natbib}
\usepackage{url}
\graphicspath{{images/}}
\usepackage{float}
\usepackage{blindtext}
\usepackage[justification=centering]{caption}
\usepackage{SCITEPRESS}     

\begin{document}

\title{Function-as-a-Service Benchmarking Framework}

\author{\authorname{Roland Pellegrini\sup{1}, Igor Ivkic\sup{2,3} and Markus Tauber\sup{3}}
\affiliation{\sup{1}Plan-B IT, Vienna, Austria}
\affiliation{\sup{2}Lancaster University, Lancaster, UK}
\affiliation{\sup{3}University of Applied Sciences Burgenland, Eisenstadt, Austria}
\email{roland.pellegrini@plan-b.at, i.ivkic@lancaster.ac.uk, markus.tauber@fh-burgenland.at}
}
\vspace{-5cm}

\abstract{Cloud Service Providers deliver their products in form of ''as-a-Service'', which are typically categorized by the level of abstraction. This approach hides the implementation details and shows only functionality to the user. However, the problem is that it is hard to measure the performance of Cloud services, because they behave like black boxes. Especially with Function-as-a-Service it is even more difficult because it completely hides server and infrastructure management from users by design. Cloud Service Prodivers usually restrict the maximum size of code, memory and runtime of Cloud Functions. Nevertheless, users need clarification if more ressources are needed to deliver services in high quality. In this regard, we present the architectural design of a new Function-as-a-Service benchmarking tool, which allows users to evaluate the performance of Cloud Functions. Furthermore, the capabilities of the framework are tested on an isolated platform with a specific workload. The results show that users are able to get insights into Function-as-a-Service environments. This, in turn, allows users to identify factors which may slow down or speed up the performance of Cloud Functions.}
 
\onecolumn \maketitle \normalsize \vfill

\section{\uppercase{Introduction}}
\label{sec:introduction}

\begin{figure}[ht]
	\centering
	\includegraphics[width=0.49\textwidth]{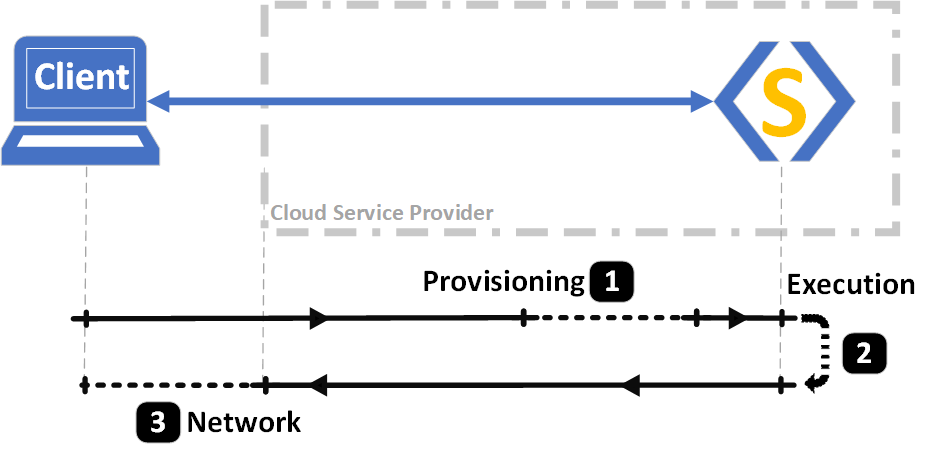}
	\caption{Various performance issues (1, 2, 3) while invoking and executing a Cloud service (S)}
	\label{fig:faas}
\end{figure}

\noindent In the past, many benchmark tools have been developed and characterized to evaluate the performance of hard- and software components \citep{berry_scientific_1991}. The standardization of those benchmark tools by organizations such as Standard Performance Evaluation Corporation \citep{spec2019} made it possible to objectively assess single components or an entire system. Traditionally, computer benchmarks are mainly focused on performance qualities of computer systems, which are under the direct control of the benchmarking program. Cloud service benchmarking, in contrast, is about IT benchmarking of software services where the underlying infrastructure is abstracted away \citep{bermbach2017}. Nevertheless, it can not be avoided that Cloud services may become unavailable, slow to respond, or limited with regards to resources such as memory, computing time, or maximum number of requests. Therefore, Cloud Service Consumers (CSC) need a clear statement if a Cloud service fulfills its purpose  with the existing resources or if additional Cloud resources are needed. Consequently, new methods for benchmarking and performance metrics for Cloud services must be taken into account. As shown in Figure \ref{fig:faas}, an external client calls a Cloud service (S), which is provided by a Cloud Service Provider (CSP). Since the client and Cloud service are geographically dispersed, a variety of performance issues can occur, which may be a compound result of many various causes. In this scenario, the request may be delayed by service provisioning mechanism (1) inside the CSP infrastructure, while code execution (2) of the Cloud Service itself, or by network delays (3) between CSP and the client. Since the Cloud infrastructure is hidden by design, the CSC is not able to identify the exact location of the root cause. As a result, the CSC can not get to the bottom of the issue in order to solve the problem as well as to prevent it in the future.

Function-as-a-Service (FaaS) is a relatively young technology in the field of Cloud Computing, which provides a platform for CSC to develop and run applications without managing servers and providing low-level infrastructure resources. In a FaaS environment, applications are broken down into granular function units of scale, which are invoked through network protocols with explicitly specifed or externally triggered messages \citep{spillnersnafu2017}. All functions are executed in a fully-managed environment where a CSP handles the underlying infrastructure resources dynamically and manages the runtime environment for code execution. The software development and application management remains with the CSC whereas the responsibility for running and maintaining the environment shifts from the CSC to the CSP. 

In recent years, a large number of commercial and Open-source FaaS platforms have been developed and deployed. Under the hood, all platform are heterogeneous in nature, because they use different hardware and software components, different runtime systems and programming languages, routing gateways, resource management tools and monitoring systems. As an example, \cite{malawski2017} points out that most platforms use Linux, but Azure functions run on Windows. With the rising popularity of FaaS, an increasing need to benchmark different aspects of FaaS platforms has been recognized by research and industry \citep{kuhlenkamp2018}. However, the authors add that it remains challenging to efficiently identify a suitable benchmarking approach. They furthermore point out that there is a need for efficiently identifying the current state of the art of experiments, which validates FaaS platforms.

In this paper, we present a prototype of a framework for benchmarking FaaS, which takes Cloud Functions and the underlying environment into account. We describe the architectual design, the components, and how they interact with each other. Typically, benchmark tools are used to evaluate the performance of hard- or software components in terms of speed or bandwidth. However, we show how this benchmarking framework can also be used to identify limitations and restrictions on top of the FaaS infrastructure. This, in turn, helps CSC to identify if the overall performance of Cloud Functions and FaaS platform meets their business requirement. 

The remainder of this paper is as follows: we  provide a summary of the related work in Section II, followed by a detailed description of the benchmark framework archictechture in Section III. Afterwards, we present our test scenario and the results in Section IV. Finally, we conclude our work with a short summary and future work in Section V.

\vspace{-0.5cm}
\section{\uppercase{Related Work}}
\label{sec:relatedwork}

\noindent An initial introduction and guideline have been done by \cite{bermbach2017} who are mainly interested in client-observable characteristics of Cloud Services. The authors cover all aspects of Cloud service benchmarking including motivation, benchmarking design and execution, and the use of the results. The authors point out that Cloud benchmarking is important as applications depends more and more on Cloud services. However, this work describes a general picture of Cloud benchmarking but does not focus on FaaS, in particular, and specific characteristics of this new Cloud Service.

\cite{kuhlenkamp2018} identify the need to benchmark different qualities and features of FaaS platforms and present a set of benchmarking approaches. They present a preliminary results for a systematic literature review in support of benchmarking FaaS platforms. The results show that no standardized and industry-wide Benchmark suite exists for measuring the performance and capabilities of FaaS implementations. Their results indicate a lack of benchmarks that observe functions not in an isolated but in a shared environment of a Cloud Service. 

\cite{spillnerfaaster2017} analyze several resource-intensive tasks in terms of comparing FaaS models with conventional monolithic algorithms. The authors conduct several experiments and compare the performance and other resource-related characteristics. The results demonstrate that solutions for scientific and high-performance computing can be realized by Cloud Functions. The authors mainly focus on computing intensive tasks (e.g. face detection, calculation of $\pi$, etc.) in the domain of scientific and high-performance computing but they do not take other FaaS qualities of interest into account. For example, FaaS environments offer a timeout parameter, which specifies the maximum time for code execution. After this time, any code execution stops and the FaaS platform returns an error status. A timeout might become crucial when it comes to Cloud Function pricing since CSC is charged for the entire time that the Cloud Function executes. Therefore, a lack of precision can lead to early and misleading timeouts during code execution, which, in turn, can end in uncompleted tasks and high costs.

\cite{mcgrath2017} present a performance-oriented serverless computing platform to study serverless implementation considerations. Their prototype provides important insights how functions are managed and executed in serverless environments. The authors also discuss several implementation challenges such as function scaling and container discovery in detail. Furthermore, they propose useful metric to evalute the performance of serverless platform. The presented prototype and its internal architecture, in particular, helped us to design and implement the FaaS Benchmarking Framework.

\cite{malawski2017} present an approach for performance evaluation of Cloud functions and take the heterogeneity aspects into account. For this purpose, the authors developed a framework with two suites of computing-intensive benchmarks for performance evaluation of Cloud functions. Their results show the heterogeneity of Cloud Function Providers, the relation between function size and performance and how providers interpret the resource allocation policies differently. The authors conclude that there is need of research that should analyse the impact of parallelism, delays, and warm-up on performance.

\cite{lloyd2018} study the factors that influence the performance of microservices provided by serverless platforms. In detail, the authors focus on infrastructure elasticity, load balancing, provisioning, infrastructure retention, and memory reservation. For this purpose, \cite{lloyd2018} implement two dedicated functions, which are executed on the platforms Azure Functions \citep{azurefunctions2019} and AWS Lambda \citep{awslambda2019}. This approach is useful when comparing the performance of different FaaS platforms but it does not take the performance of business-related Cloud Functions into account. Therefore, the FaaS Benchmark Framework presented in this paper is able to benchmark Cloud Functions which are not specifically adapted for benchmarking. Instead, it is also applicable for performance evaluation of production-related Cloud Functions including the underlying FaaS platforms.

\cite{hwang2016} present a generic Cloud Performance Model and provide a summary of useful Cloud Performance Metrics (CPM) on three levels: Basic performance metrics, Cloud capabilities, and Cloud productivity. The Basic performance metrics include traditional metrics such as execution time or speed. The Cloud capabilities describe throughput, bandwidth, and network latency. Finally, Cloud productivity deals with productivity metrics such as Quality of Service (QoS), Service Level Agreement (SLA) and security. The authors encourage the Cloud community to test Cloud capability in big-data analytics and machine learning intelligence. In particular, they argue that the Cloud community is short of benchmarking tests.  In this regard, the authors motivated us to develop the FaaS Benchmarking Framework and to adapt existing tests for FaaS platforms. This includes also the analysis and feasibility of appropriate FaaS performance metrics.

\cite{back2018using} discuss the use of a microbenchmark in order to evaluate how different FaaS solutions behave in terms of performance and cost. For this purpose, the authors develop a microbenchmark in  order to  investigate the observable behavior with respect to the computer/memory relation of different FaaS platforms, and the pricing models currently being in use.

\cite{mohanty2018} analyse the status of Open-source serverless computing frameworks \cite{fission2019}, \cite{kubeless2019} and \cite{openfaas2019}. For this purpose, the authors evaluate the performance of the response time and ratio of successfully responses under different loads. The results show that Kubeless has the most consistent performance across different scenarios. In constrast, the authors notice that OpenFaaS has the most flexible architecture with support for multiple container orchestrators. 

Finally, the FaaS Benchmark Framework presented in this paper is built upon a previous work of \cite{pellegrini2018}, who present an initial investigation of benchmarking FaaS and outline the architectural design. \cite{pellegrini2018} present a two-tier architecture of a benchmarking framework where requests are invoked by a sender and processed by a Cloud Function on the CSP platform. This paper takes up the idea of previous research and introduces an additional third component that enables testers to evaluate the performance of the FaaS platform more precisely. Details about this third component and its improvements for benchmarking FaaS are discussed in Section III.
\vspace{-0.1cm}

\section{\uppercase{FaaS Benchmarking Framework}}
\label{sec:architecture}

\begin{figure}[h]
	\centering
	\includegraphics[width=0.45\textwidth]{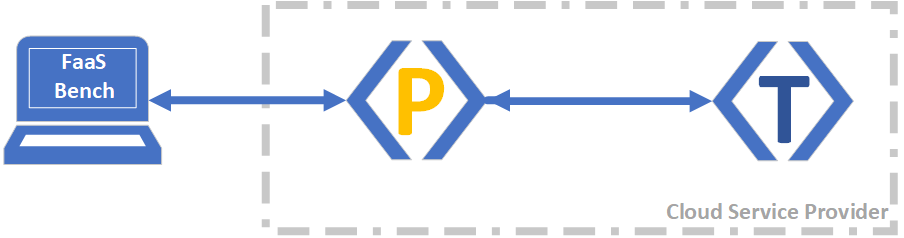}
	\caption{Architecture of the Faas Benchmarking Framework with FaaSBench, PCF (P), and TCF (T)}
	\label{fig:architecture}
\end{figure}

\noindent The components of the FaaS Benchmarking Framework are shown in Figure \ref{fig:architecture}. Basically, it consists of two software components, a Java-based application (FaaSBench) and a JavaScript-based Proxy Cloud Function (PCF). The FaaSBench is responsible to create and invoke workloads on the Target Cloud Function (TCF) while the PCF collects benchmark relevant metrics of the TCF and the FaaS platform.

\vspace{0.2cm}

The following subsections provide a detail description about the architecture of the FaaS Benchmark Framework, the components involved and how they communicate and exchange data with each other.

\subsection{Framework Architecture}
\label{sec:architecture2}

As illustrated in Figure \ref{fig:architecture}, the proposed architecture divides the intial request into two separate calls. The first call between FaaSBench and the PCF allows the measurement of transmission relevant metrics such as bandwith, byte size, or transfer rate. Since Cloud Functions are invoked over network protocols, latencies and delays during the data transmission can be easily identified. In contrast, the second call between PCF and TCF is used to measure benchmark relevant metrics on top of the FaaS platform. This includes not only the time measurement of compute-intensive tasks on top of the CSP's infrastructure but also to measure the routing time for invocation calls  the time needed for launching a new function instance, and the time a function instance stays active.

In addition, the proposed architecture supports a variety of implementation and application options. Since the PCF is under the control of the CSC, it can be easily adapted, modified, and extended with additional benchmark-relevant profiling methods or logging functionalities without adapting the source code of the TCF. Furthermore, the PCF can decode and analyze the communication between any client and the TCF in both, production and test environments. For this purpose, the PCF can split, enrich and invoke new requests on-demand, if necessary. In summary, the introduction of the PCF allows the measurement of time and resource critical factors of Cloud Functions on top of FaaS platforms.

\subsection{FaaSBench application}
\label{sec:architecture_faasbench}

\noindent The first component of the FaaS Benchmark Framework is a Java-based application named FaaSBench. This tool is responsible for generating workloads, invoking requests, collecting metrics, and providing statistical reports. Since the FaaSBench application is written in Java, it can be installed on any client machine which provides a Java runtime environment. 

Technically, the FaaSBench application consists of a set of Java classes, which are grouped in three main packages. The first package, the Generator, consists of Java classes, which are responsible for executing a set of benchmark tests by extracting workloads from the properties. For this purpose, the Generator supports different operations, which take technical aspects and specific characteristics of FaaS platforms into account. In the context of this paper, the following three types of operations have been implemented: first, if the workload focuses on peak performance evaluation, the Generator executes requests or in groups of batches, synchronously or asyncronously. Second, in order to test the responsiveness of the FaaS environment, the Generator uses an operation with an automatic retry logic, which progressively longer waits between the retries. This operation evaluates the time when functions need to be provisioned in as a short time span due to incomming requests. Finally, if the maximum time for Function code execution is crucial, the Generator executes an operation, which provokes a timeout error, either on the FaaS Gateway or on Cloud Function level. Typical uses are therefore situations when it comes to resource planing in terms of timing and accuracy.

\begin{figure}[h]
	\centering
	\includegraphics[width=0.48\textwidth]{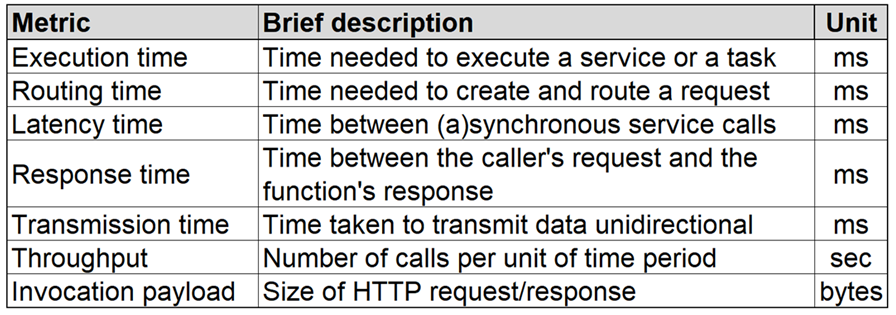}
	\captionof{table}{Metrics provided by the\\FaaS Benchmarking Framework}{}
	\label{fig:metrics}
\end{figure}

The purpose of the second package, the Metrics, is three-fold: first, it records the status of each invocation call. This logic helps to identify and distinguish successful invocation calls from unsuccessful or uncompleted calls. Second, it records the responses of the TCF for the purpose of transparency and providing proof of the correctness. Finally, the package also records all benchmark relevant performance metrics during each benchmark run. At the time of writing, the FaaS Benchmark Framework records only a set of well-established metrics. A detailed overview of all supported metrics are listed in Table \ref{fig:metrics}, including a brief description and the measured units, expressed  in milliseconds (ms), seconds (sec), or in bytes.

Finally, the Java-classes of the Reporting package provide different forms of presentation of the measured values (e.g. total runtime, total byte size, etc.). The package generates a set of logfiles and comma-separated values  (CSV) files, which can be processed by a various spreadsheet applications or command-line programs for data visualization.

\subsection{Proxy Cloud Function (PCF)}
\label{sec:architecture_pcf}

\noindent The second and innovative component of the FaaS Benchmark Framework is the PCF, a JavaScript-based Cloud Function, which is installed on a FaaS platform. From the viewpoint of a workflow, the PCF is placed between the FaaSBench application and the TCF. The PCF receives requests from the FaaSBench application and forwards them to the TCF. Afterwards, it sends all responses of the TCF back to the FaaSBench application. 

Since the PCF acts as an intermediary, it gains new opportunities and solutions by solving following problems efficiently: 

\begin{enumerate}
	
	\item The PCF is useful when any kind of modification of the testbed for benchmark-relevant profiling integration is not allowed or possible. In addition, the PCF can be easily adapted and modified to meet very special requirements of the TCF. For example, a request can be split up and its data can be enriched with additional workflow-relevant information at runtime. As a result, the PCF can be used not only in test environments but also in production-related environments.
	
	\item The PCF is able to collect useful performance metrics of the FaaS environment itself. As an intermediary, the PCF is able to measure and evaluate the routing time of invocation calls, the throughput and the data transfer capabilities of the Cloud network inside the FaaS environment. 
	
	\item The PCF is useful when problems with clock synchronization between the FaaSBench and the TCF occur. Since the CSP dictates the system time for the FaaS platform, PCF and TCF can rely on this unifom time basis.
	
	\item Multiple PCF instances can be installed in different regions of the CSP. In terms of \cite{google2019}, a region is defined as a geographical location where resources and services are hosted and executed.  Running multiple PCF instances in different regions can be helpful in choosing a location that is close to the service and has minimal network and transmission latency.
	
\end{enumerate}

\subsection{Target Cloud Function (TCF)}
\label{sec:architecture_tcf}

\noindent TCFs usually represent a set of Cloud Function, which are optimized for production environments. They are characterized by the fact that code modifications for a benchmark-relevant profiling integration are not permitted or possible. Typical use-cases for productive TCF are data and event processing, server-side back-ends for mobile apps, or the orchestration of microservice workloads. The FaaS Benchmark Framework uses representative real-world workloads for performance evaluation of productive TCFs.

However, the FaaS Benchmark Framework also includes a set of non-productive TCFs. These functions include benchmark-relevant profiling methods and execute synthetical workloads, which take the unique characteristic of FaaS platforms into account. For example, a TCF may provoke a timeout errors when executing a long-running operation for determining the behavior of the FaaS platform. Another non-productive TCF supports workloads for evaluating the responsivness of FaaS platform by executing requests periodically with progressively longer waits between two requests.

By using productive and non-productive TCFs, the FaaS Benchmark Framework can be used in test and production-related environments without the need of implementing a complex testbed. Due to performance reasons, however, all non-productive TCFs do not log information to external logging tools. Instead, all benchmark-relevant data are kept in memory-based messages and distributed between the components of the FaaS Benchmarking Framework.

\subsection{Communication flow}
\label{sec:architecture_communication}

\begin{figure}[h]
	\centering
	\includegraphics[width=0.5\textwidth]{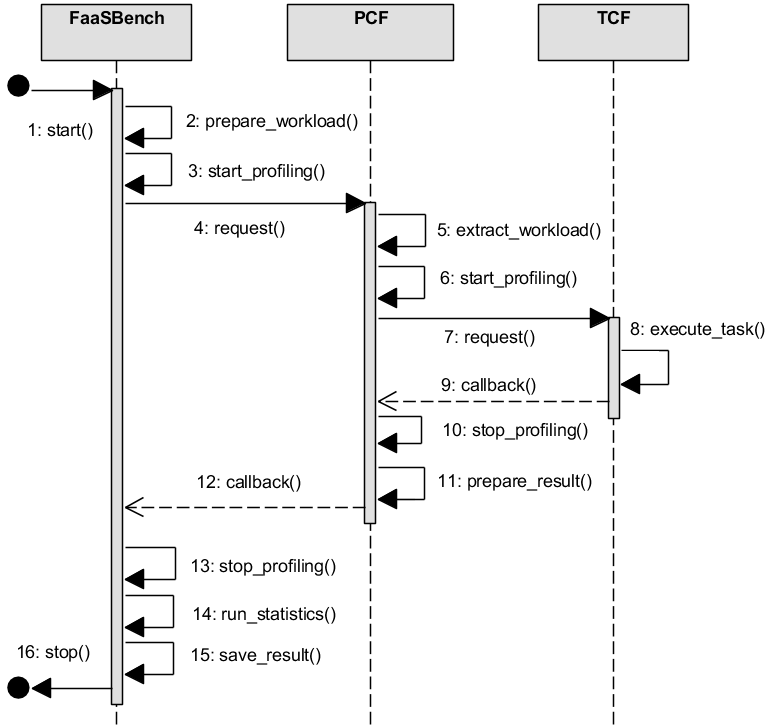}
	\caption{Interaction between the components\\FaaSBench, PCF, and TCF}
	\label{fig:sequence}
\end{figure}

\noindent The interaction between the FaaSBench, the PCF, and the TCF in sequential order as well as their corresponding method calls are shown in Figure \ref{fig:sequence}. 

The communication between the components is as follows: After starting (1) the FaaSBench application, the program prepares the workload (2), starts the corresponding profiling method (3), and invokes the initial request (4). On the recipients' side, the PCF extracts the workload from the request (5), starts its internal profiling method (6), and invokes a separate request (7) to the TCF. After executing the requested task (8) and receiving (9) the result from the TCF, the PCF stops its internal profiling method (10), prepares the result message (11) and returns it back (12) to the FaaSBench application. Finally, the FaaSBench stops its internal profiling (13) as well, runs the statistical analysis (14) and saves the results locally (15). This also marks the end of the benchmark run (16).

\subsection{Data exchange}
\label{sec:architecture_data_exchange}

\noindent FaaSBench, PCF, and TCF (if non-productive) use a common way to exchange workloads, data and logging information. Both components communicate with each other by using JSON-based (JavaScript Object Notation) messages, which contain self-describing data fields. Each field name starts with a prefix, which indicates the location of its origin. For example, the field ''proxy\_start\_time'' refers to the internal timer of the PCF, and only the PCF is able to create this field.

In order to evaluate the performance  of a TCF for counting letters, the following example illustrates a JSON message used for the communication between FaasBench and the PCF.

\begin{small}
	\begin{verbatim}
	{ "faasbench_workload_uuid": "112c338d",
	"target_uri": "https://faas:8080/func/word",
	"faasbench_workload_data":"F a a S" }
	\end{verbatim}
\end{small}

The request message is structured as follows: the field ''faasbench\_workload\_uuid'' specifies the universally unique identifier (UUID) for this request. This ensures uniqueness over the entire lifetime of the benchmark run. The field ''target\_uri'' refers to the location of the TCF and instructs the PCF to forward this request to it. Finally, the key ''faasbench\_workload\_data'' contains the workload which has to be processed by the TCF. In the example above, the workload consists of a text with four (4) letters, which has to be counted by the TCF.

After the TCF has executed its task, the following sample response shows an extract of the JSON message created by the PCF and reported back to the FaaSBench application. 

\begin{small}
	\begin{verbatim}
	{ "proxy_workload_uuid": "112c338d",
	"target_start_time":1533675892375,   
	"target_stop_time":1533675892401,
	"target_run_time_hr_seconds":0,
	"target_run_time_hr_nanoseconds":26250589,
	"target_workload_result": "4" }
	\end{verbatim}
\end{small}

Each response message is structured as follows: The field ''proxy\_workload\_uuid'' specifies the UUID of this response which must be identical to the original UUID of ''faasbench\_workload\_uuid'', otherwise the workload will be marked as invalid. The fields ''target\_start\_time'' and ''target\_stop\_time'' refer to the start and stop time of the TCF. All time values are representatives of the Long data type, and refer to the Unix Epoch Time \citep{theopengroup2019}. The fields ''target\_run\_time\_hr\_seconds'' and ''target\_run\_time\_hr\_nanoseconds'' refer to the time of TCF execution in high-resolution realtime of a [second, nanosecond] tuple array. Finally, the result of the function execution can be found in the field ''target\_workload\_result''.

\vspace{-0.5cm}

\section{\uppercase{Evaluation scenario}}
\label{sec:scenario}
\noindent The last section described the architecture of the FaaS Bechmark Framework. This section demonstrates how the framework is used on basis of a concrete example and discusses the results of the benchmark run.

\subsection{Setup \& Procedure}

\noindent The testbed is based upon the Open-source FaaS implementation \cite{openfaas2019}, a framework for building serverless functions on top of containers. The testbed consists of a virtual machine (VM) running Debian 9.1 with 4 GigaByte (GB) of memory allocated, which is deployed on a physical QNAP server with a quad-core Intel i5-4590S CPU @ 3.00 GigaHertz (GHz) and 16 GB of memory in total. This isolated testbed eliminates any effects of interference and perturberation that could affect the experiment. A PCF and a non-productive TCF, as discussed in Section \ref{sec:architecture_pcf} and \ref{sec:architecture_tcf}, are deployed on the same VM. The experiment does not focus on performing high-performance computating tasks, which are preferably used in most researches. Instead, the proposed testbed allows the study of request and response messages on top of the FaaS platform. For this purpose, the TCF expects a text and answers immediately with the number of words and letters in this text. In the context of the work, a word represents a character separated by at least one space from another character.

\begin{figure}[ht]
	\centering
	\includegraphics[width=0.46\textwidth]{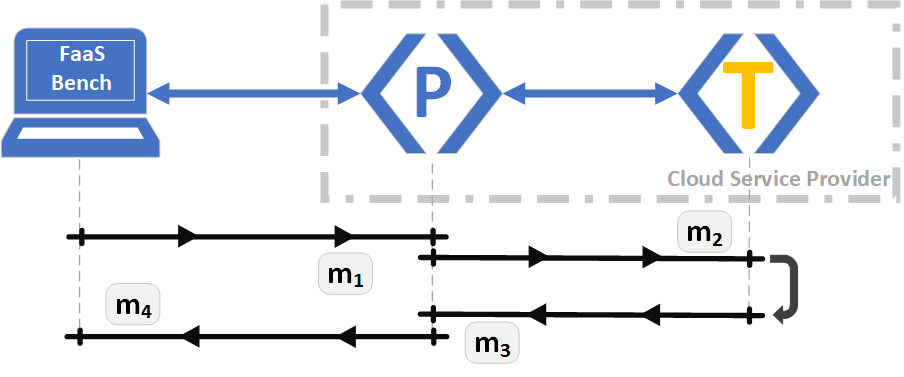}
	\caption{Measurement points with routes for HTTP traffic}
	\label{fig:workflow}
\end{figure}

A schematic overview of the workflow is shown in Figure \ref{fig:workflow}, illustrating four measurement points (m1, m2, m3, and m4) where HTTP traffic is decoded, stored, and prepared for further analysis. 

The aim of the benchmark run is to evaluate the size of the Hyptertext Transfer Protocol (HTTP) requests/responses inside the FaaS environment in relation to the size of the workload. For this purpose, the benchmark run uses different workloads (10, 10\textsuperscript{2}, 10\textsuperscript{3}, 10\textsuperscript{4}, 10\textsuperscript{5}, and 10\textsuperscript{6} words). The first measurement point (m1) evaluates the size of the initial request, sent by FaaSBench. Measurement point m2 examines the request size of the HTTP invocation for the TCF, while measurement point m3  evaluates the size of the reply message. Finally, measurement point m4 collects the size of the response message, returned by the PCF. In addition, more data such as runtimes or latencies are measured by FaaSBench, PCF, and TCF in order to identify variances and trends. 

\subsection{Results \& Findings}

\begin{figure}[ht]
	\centering
	\includegraphics[width=0.48\textwidth]{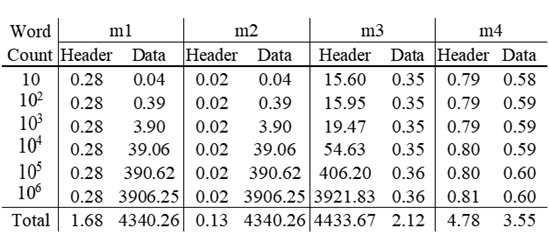}
	\captionof{table}{Size of HTTP messages per measurement point, all values in KB}{}
	\label{fig:measurement1}
\end{figure}

\noindent Table \ref{fig:measurement1} illustrates the collected measurement values for progressively increasing word count per row. Each column of the table refers to a measurement point (m1, m2, m3, m4) that evaluates the size of the HTTP message header (Header) and the message body data (Data) in Kilobyte (KB). Finally, the last row of the table summarizes the total size in KB.

Measurement points m1, m2 and m4 report an increase of the size of message body data in sync with the payload, while the size of the message headers remains stable. However, measurement point m3 shows a completely different picture. While the size of the HTTP message body data increases slowly in sync with the result of the word count, the HTTP message header increases rapidly. Analyses of the logging data (see \ref{sec:architecture_data_exchange} for more details) demonstrate that the reason for the growth of the header is caused by the underlying JavaScript runtime environment. In detail, the engine creates an internal message object, which contains the result of the function call but also the original workload data of the benchmark run. As a result, the size of the HTTP message header increases in sync with the workload. This, in turn, may have impact to total runtime caused by the amount of time it takes for the HTTP response to travel from TCF to PCF. 

\begin{figure}[ht]
	\centering
	\includegraphics[width=0.47\textwidth]{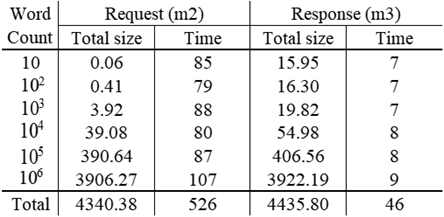}
	\captionof{table}{Time for transmission (in ms) of HTTP requests and responses (in KB) per measurement point m2 and m3}{}
	\label{fig:measurement2}
\end{figure}

In order to get a clear picture, Table \ref{fig:measurement2} summarizes the effect to the transmission time in response to the HTTP requests and responses. The column "Request (m2)" refers to the route PCF to TCF and represents the total byte size (in KB) of the HTTP requests and the transmission time (in ms) per HTTP request and word count. In contrast, the column "Response (m3)" refers to the route TCF to PCF and represents the  total byte size (in KB) of the HTTP responses and the transmission time (in ms). The last row summarizes the total byte size of the HTTP requests/responses sent and their total runtimes.

As expected, the transmission times for HTTP responses increases in sync with the word count but not in the same way as the transmission times for HTTP requests. The time values for HTTP requests range from 80 up to 107 ms, with a total runtime of 526 ms. However, the time values for HTTP responses range from 7 up to 9 ms, resulting in a total runtime of 46 ms. The results are even more surprising because the total byte size of HTTP requests are only slightly different from the total byte size of the corresponding HTTP responses. A possible explanation for this is the use of caching mechanism inside the \cite{nodejs2019} runtime. However, a detailed examination of the reasons for this divergence is not in focus of this paper but offers opportunities for further research in this area. 

\subsection{Discussion}

\noindent The FaaS Benchmark Framework demonstrates that CSCs are able to get insights into the FaaS environment. It overcomes the limitation of abstraction because the PCF acts as an middleman between the FaaSBench and a TCF. When the FaaSBench sends a request to our PCF, it forwards the request to the TCF and waits for the response. As an intermediary, it monitors the HTTP traffic and measures the byte size of outbound requests and inbound responses. This approach allows CSC to compare the size of all inbound and outbound HTTP messages, which are handled by the FaaS environment. As demonstrated in the experiment, the FaaS Benchmarking Framework uncovers HTTP overhead while transmitting data inside the FaaS platform.

In this experiment, we have evaluated and analyzed the size of the HTTP requests/responses inside the FaaS environmnent in relation to the size of the workload. We have shown that the cascading of Cloud Functions generates an overhead in the HTTP message header with the potential of negative performance impacts for the runtime environment. However, we have not evaluated critically the performance of the testbed. The reason for this is that most research, as discussed in Section \ref{sec:relatedwork}, have evaluated and compared the performance of commercial and Open-source FaaS platform. Instead, we believe that the performance evaluation of a business-related Cloud Function with its underlying FaaS platform is a more reliable indicator for the CSC.

We encountered three issues while preparing and executing the experiment. First, we noticed that time synchronization becomes crucial when it comes to benchmarking, monitoring and realtime control of Cloud Functions as timestamps must by synchronized across multiple geographical regions, between different Cloud infrastructures, and clients. Second, we observed performance degradation caused by logging services of the FaaS platform. This is because logging requires additional resources that must be handled by the FaaS platform during the function's execution. Finally, we observed issues with performance isolation between the coresident PCF and TCF instances. The phenomenon with ''noisy neighbors'' may be fixed by isolation through dedicated FaaS platforms or by moving workloads across physical servers. We therefore encourage researchers to work jointly in developing methods for dealing with these issues. 

The proposed architecture has room for improvements. First, the FaaS Benchmark Framework has been tested with different sizes of workloads. However, the limiting factor are not the number of requests or the size of workloads but computer memory and network bandwidth. The FaaS Benchmark Framework code is not optimized for benchmark runs that exceed these limits. Second, the PCF does not provide any functionality to store a specific workload temporary on the FaaS Platform when re-using it for multiple runs. Finally, CSPs usually provide logging functionality which records detailed activities of the FaaS Platform. Currently, FaaSBench and PCF are not able to retrieve logging data from FaaS platforms.

\section{\uppercase{Conclusion}}
\label{sec:conclusion}

\noindent In this paper, we introduced a benchmarking framework for measuring the performance of Cloud Functions and FaaS environments. First, we explained the idea behind FaaS, its benefits for CSC and CSP, but also the challenges for benchmarking FaaS. Next, we explained the architecture of the framework, its components, and how they are linked together. Finally, in Section IV we prepared and executed a benchmark run executed on an isolated FaaS platform. The results showed a growth of message sizes generated by the underlying runtime environment and identified the root cause. In summary, the main contribution of this paper is a FaaS Benchmarking Framework framework which allows performance evaluation of FaaS environments by a using proxy-based implementation.

The current version of the FaaS Benchmark Framework provides interesting insights of a FaaS environment. Nevertheless, many experiments, adaptations, and scientific intensification of know\-ledge in FaaS have been left for the future. First of all, we plan to migrate our project to OpenJDK \citep{openjdk2019}. Afterwards, we will make the soure code of the FaaS Benchmark Framework publicly available. Furthermore, we plan to implement additional PCF and non-productive TCFs written in different programming languages and supporting more runtime environments. In addition, we plan to define new workloads, which are specially tailored to the features of FaaS enviromnents. For example, we want to examine how precisely a FaaS environment works in case of time controlling for code execution versus billing. For this purpose, we plan to examine the timeout parameter of Cloud Functions, which specifies the maximum time for code execution. Inaccuracies in the timing may lead to uncompleted tasks and high costs for CSCs. More workloads for high-performance computing will complete the portfolio. 

Finally, we consider using the FaaS Benchmarking Framework on public FaaS platforms. For future work, it would be interesting to know how the benchmarking results of the different FaaS providers would differ from the reference testbed. We also plan to analyse different configurations of existing autoscaling solutions in respect to performance.

\section*{\uppercase{Acknowledgements}}
\noindent The research has been carried out in the context of the project MIT 4.0 (FE02), funded by IWB-EFRE 2014-2020. The technical infrastructure needed for the research has been provided by Plan-B IT.

\vfill
\bibliographystyle{apalike}
{\small
\bibliography{references}}

\vfill
\end{document}